\documentclass[twocolumn,showpacs]{revtex4}  
 \usepackage[latin1]{inputenc}
 \usepackage{amsmath}
 \usepackage{epsfig}
 \usepackage{subfigure}
 \usepackage{placeins}
 \usepackage{floatflt}
 \usepackage{amsmath}
 \usepackage{amssymb}
 \usepackage{wrapfig}
 \usepackage[]{tocbibind}
 \usepackage{graphicx}

 \usepackage{ifthen}
 \usepackage{times}

\makeatletter
\newcommand\figcaption{\def\@captype{figure}\caption}
\newcommand\tabcaption{\def\@captype{table}\caption}
\makeatother

\newlength{\mylinewidth}
\begin{document}


\title{ $\mathbf{\phi}$ decay: an relevant source for ${K^-}$ production at SIS energies?}
\author{\footnotetext{corresponding authors:}  G.~Agakishiev$^{8}$, A.~Balanda$^{3,d}$, B.~Bannier$^{5}$, R.~Bassini$^{9}$,
D.~Belver$^{15}$, A.V.~Belyaev$^{6}$, A.~Blanco$^{2}$, M.~B\"{o}hmer$^{11}$, J.~L.~Boyard$^{13}$,
P.~Braun-Munzinger$^{4}$
% ...
}

\affiliation{
\footnotesize
\begin{center}
(HADES collaboration)
\end{center}
\\\mbox{$^{1}$Istituto Nazionale di Fisica Nucleare - Laboratori Nazionali del Sud, 95125~Catania, Italy}\\
\mbox{$^{2}$LIP-Laborat\'{o}rio de Instrumenta\c{c}\~{a}o e F\'{\i}sica Experimental de Part\'{\i}culas , 3004-516~Coimbra, Portugal}\\
\mbox{$^{3}$Smoluchowski Institute of Physics, Jagiellonian University of Cracow, 30-059~Krak\'{o}w, Poland}\\
\mbox{$^{4}$GSI Helmholtzzentrum f\"{u}r Schwerionenforschung GmbH, 64291~Darmstadt, Germany}\\
\mbox{$^{5}$Institut f\"{u}r Strahlenphysik, Forschungszentrum Dresden-Rossendorf, 01314~Dresden, Germany}\\
\mbox{$^{6}$Joint Institute of Nuclear Research, 141980~Dubna, Russia}\\
\mbox{$^{7}$Institut f\"{u}r Kernphysik, Johann Wolfgang Goethe-Universit\"{a}t, 60438 ~Frankfurt, Germany}\\
\mbox{$^{8}$II.Physikalisches Institut, Justus Liebig Universit\"{a}t Giessen, 35392~Giessen, Germany}\\
\mbox{$^{9}$Istituto Nazionale di Fisica Nucleare, Sezione di Milano, 20133~Milano, Italy}\\
\mbox{$^{10}$Institute for Nuclear Research, Russian Academy of Science, 117312~Moscow, Russia}\\
\mbox{$^{11}$Physik Department E12, Technische Universit\"{a}t M\"{u}nchen, 85748~M\"{u}nchen, Germany}\\
\mbox{$^{12}$Department of Physics, University of Cyprus, 1678~Nicosia, Cyprus}\\
\mbox{$^{13}$Institut de Physique Nucl\'{e}aire (UMR 8608), CNRS/IN2P3 - Universit\'{e} Paris Sud, F-91406~Orsay Cedex, France}\\
\mbox{$^{14}$Nuclear Physics Institute, Academy of Sciences of Czech Republic, 25068~Rez, Czech Republic}\\
\mbox{$^{15}$Departamento de F\'{\i}sica de Part\'{\i}culas, Univ. de Santiago de Compostela, 15706~Santiago de Compostela, Spain}\\
\mbox{$^{16}$Instituto de F\'{\i}sica Corpuscular, Universidad de Valencia-CSIC, 46971~Valencia, Spain}\\
\\
\mbox{$^{a}$ also at ISEC Coimbra, ~Coimbra, Portugal}\\
\mbox{$^{b}$ also at Technische Universit\"{a}t Dresden, 01062~Dresden, Germany}\\
\mbox{$^{c}$ also at Dipartimento di Fisica, Universit\`{a} di Milano, 20133~Milano, Italy}\\
\mbox{$^{d}$ also at Panstwowa Wyzsza Szkola Zawodowa , 33-300~Nowy Sacz, Poland}\\
\mbox{$^{e}$ also at Excellence Cluster Universe, Technische Universit\"{a}t M\"{u}nchen, Boltzmannstr.2, D-85748, Garching, Germany}\\
}
\date{\today}

\begin{abstract}
We present phase space distributions and multiplicities of ${K^+}$, ${K^-}$ and $\phi$ mesons
produced in Ar+KCl reactions at a kinetic beam energy of 1.756 AGeV
and measured with the HADES spectrometer.
The inverse slope parameters and yields of kaons supplement the systematics
of previous measurements. The percentage of ${K^-}$ mesons coming from $\phi$ decay
is found to be $18 \pm 7 \%$.
\end{abstract}
\maketitle

\section{Introduction}

The systematic study of $K^\pm$ sub-threshold production yields, phase space distributions
and flow observables in relativistic heavy-ion collisions at various beam energies and for
various system sizes and centralities has attracted much attention, in particular in the context of
in-medium properties of ${K^+}$ and ${K^-}$ mesons
\cite{Kaplan86,Brown94,Hartnack94,Kolomeitsev95,Waas96,Ko96,Li97,Cassing97,Schaffner97,Lutz98}.
Corresponding experiments have been the focus of studies by the KaoS
\cite{Sturm01,Foerster03,Senger04,Uhlig05,Foerster07}
and FOPI \cite{Ritman95,Best97,KWisnia00,Crochet00,Mangiaro03}
collaborations over the last two decades,
following up on the pioneering experiments at Bevalac \cite{Bevalac}.
The heavy-ion data are supplemented by proton-induced reactions on nuclei
\cite{Scheinast06,ANKE02} with measurements of inclusive ${K^+}$ and ${K^-}$ production.
Concerning the elementary proton-proton reactions, the recent measurements
by COSY-11 \cite{COSY_11}, TOF \cite{TOF} and ANKE \cite{ANKE06,ANKE08}
collaborations provided valuable reference data.

High statistics data for ${K^+}$ production allowed to draw conclusions
on the ${K^+}$-nucleon potential and on the nuclear equation of state \cite{Sturm01,Senger04}. From the analysis of the out-of-plane \cite{Shin98,Li96} and sideward flow \cite{Crochet00,Brat97}, the ${K^+}$-nucleon potential is deduced to be mildly repulsive, and the measured ${K^+}$ production yields are in agreement with the scenario of a soft nuclear equation of state (incompressibility modulus $K_N\approx 200$ MeV) \cite{Fuchs01,Hartnack02}.
The interpretation of the ${K^-}$ data is still undergoing systematic evaluations to pin down the value of the presumed attractive ${K^-}$-nucleon potential and its momentum dependence.\\
Recent data \cite{Foerster07} gave an important contribution
towards understanding the production mechanism of ${K^{\pm}}$ mesons.
Indeed, the combined analysis of ${K^+}$ and ${K^-}$ suggests
that a substantial part of the observed
${K^-}$ mesons is due to a strangeness exchange mechanism \cite{Hartnack03}.
Furthermore ${K^+}$ and ${K^-}$ exhibit distinctively different in-medium properties,
both in theoretical approaches \cite{Kaplan86,Brown94,Hartnack94,Kolomeitsev95,Waas96,Ko96,Li97,Cassing97,Schaffner97,Lutz98,Fuchs01,Hartnack02,Hartnack03,Fuchs06}
and in experiments \cite{Sturm01,Foerster03,Senger04,Foerster07,KWisnia00,Crochet00,Scheinast06}.

The spectrometer HADES \cite{hadesSpectro}, primarily designed to measure
di-electrons \cite{HADES-PRL07}, has been recently employed
for the identification of strange mesons as well, showing high purity and efficiency
for particle identification and excellent reconstruction capability
of secondary decay vertices \cite{PhD_Schmah}.
For the first time, a combined and inclusive identification of
${K^+}$, ${K^-}$ and $\phi$ mesons was carried out in the same experimental set-up at a sub-threshold beam energy.
(``Sub-threshold'' refers to free nucleon-nucleon collisions and applies here
to the ${K^-}$ and $\phi$ channels).
Our measurement allows firm conclusions on the fraction of ${K^-}$ mesons that originate from the $\phi$ decay. It was
argued in \cite{BK_Kotte} that, if a substantial part of the observed
${K^-}$ stems from $\phi$ decays, the interpretation of data in terms
of a strongly reduced effective anti-kaon mass in the nuclear medium
needs a proper account of this channel. Indeed, the data in \cite{Mangiaro03},
unfortunately hampered by poor statistics, gave a first hint to the
importance of ${K^-}$ feeding from $\phi$ decays.
Here, we report on the the phase-space distributions and yields of $K^\pm$ and $\phi$
mesons in the reaction Ar + KCl at a kinetic beam energy of 1.756 AGeV.
We provide an estimate of the feeding contribution of $\phi$ decays
to the ${K^-}$ abundance.\newline
Our paper is organized as follows. Section II describes the experiment
and section III is devoted to the data analysis. Experimental results on
transverse mass spectra, rapidity distributions and multiplicities of
$K^\pm$ and $\phi$ mesons are presented in section IV. Our findings are discussed in
section V and summarized in section VI.

\section{The Experiment}

The experiment was performed with the
{\bf H}igh {\bf A}cceptance {\bf D}i-{\bf E}lectron {\bf S}pectrometer HADES
at the heavy-ion synchrotron SIS
at GSI Helmholtzzentrum f\"ur Schwerionenforschung
in Darmstadt, Germany. A detailed description of the spectrometer is presented in
\cite{hadesSpectro}. In the following we summarize the main features of the apparatus.
HADES consists of a 6-coil toroidal magnet centered on the beam axis and
six identical detection sections located between the coils and covering
polar angles from $18^{\circ}$ to $85^{\circ}$.
In the measurement presented here, the six sectors comprised
a gaseous Ring-Imaging Cherenkov (RICH) detector, four planes of
Multi-wire Drift Chambers (MDCs) for track reconstruction, in the outermost plane, two opposite MDCs were not installed, and they have not been used for $K^\pm$ identification and two Time-of-Flight walls (TOF and TOFino), supplemented at forward polar angles
with Pre-Shower chambers. For each sector, the TOF and TOFino/Pre-Shower detectors are combined into META (Multiplicity and Electron Trigger Array) detectors.

An $^{40}_{18+}$Ar beam of $\sim 10^6$ particles/s was incident on a four-fold segmented
KCl target with a total thickness corresponding to $3.3\%$
interaction length. A fast diamond start detector located upstream
of the target was used to determine the interaction time.
The data readout was started by a first-level trigger (LVL1) decision,
requiring a observed charged-particle multiplicity $MUL \ge 18$ in the TOF/TOFino detectors,
accepting approximately 35\% of the nuclear reaction cross section.
It was followed by a second-level trigger (LVL2) requesting at least
one $e^{\pm}$ hit in the RICH detector per event.
This selection has, however, not introduced any bias on the hadron analysis. A fraction of 10\% of the LVL1 triggered events was stored regardless of the LVL2 condition. Comparison between the pure LVL1 and the LVL1+LVL2 events showed only insignificant differences in the total track multiplicity and hence the centrality of the reaction.

\section{Data analysis}

\subsection{Track reconstruction}
A reconstructed hadron track in the spectrometer is composed of inner and outer track segments in the MDCs
which define straight lines between the first two MDC planes, the last two MDC planes behind the magnet,
and a hit point in one of the META detectors. In the so-called cluster finder \cite{PhD_Markert,hadesSpectro} and track-segment fitting procedures possible trajectories through the MDC segments are calculated.
The pointing vector of the outer track segment was used for matching the META hit to the
MDC track segments, generating a track candidate. A Runge-Kutta algorithm was implemented
for calculating the momentum of each of the track candidates by solving the equation of motion inside the magnetic field region \cite{PhD_Sadovsky}. The quality of the META
hit-matching procedure and of the Runge-Kutta fitting (characterized by $\chi^2$ values) are used to create a list of ordered track candidates. To resolve matching ambiguities, only the track candidate with the smallest value of the product of the two $\chi^2$ values is declared a true track. Its components and associated track candidates are deleted from the candidate list. This procedure is repeated until no more track candidates are left in the list.

\subsection{Particle identification}
Since the RICH detector is only used to
distinguish $e^\pm$ from hadrons,
this system is not employed in the following analysis which deals exclusively
with hadron reconstruction. Hence, only the time-of-flight walls TOF and TOFino+Pre-Shower (this detector is used to refine the position of the TOFino hits) and the MDCs are used to identify hadrons.

\begin{figure}[!htb]
\begin{center}
\includegraphics[viewport= 5 13 536 310,angle=0,scale=0.950]{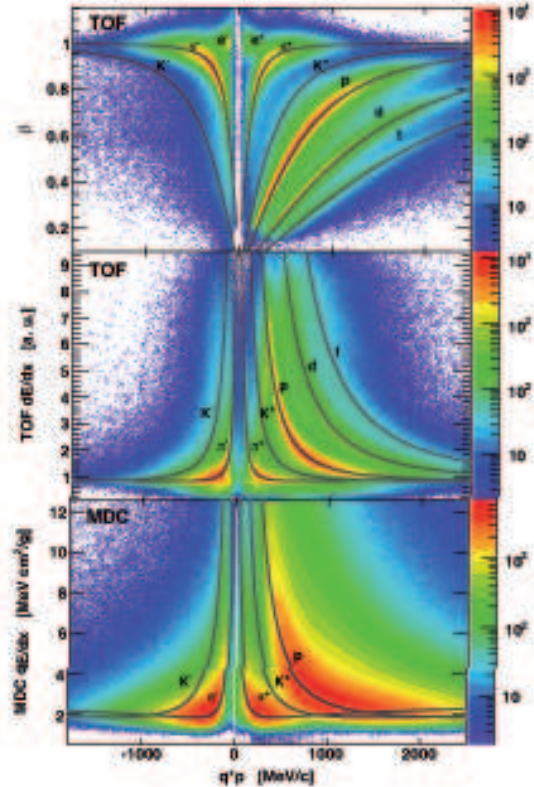}
\caption[]{Color online. Particle velocity $\beta$ (top, for the TOF detector) and $dE / dx$ (middle for TOF detector, bottom for the MDC) as a function of the momentum times the particle's polarity sign for the TOF detector.}
\label{PID_all_V3}
\end{center}
\end{figure}

The particles can be distinguished by their velocity and by their energy loss ($dE/dx$).
The particle velocity $\beta$ (in units of velocity of light $c$)
is determined by the time-of-flight
between the diamond start detector and the corresponding hit in one of the two
time-of-flight detectors and the reconstructed flight path. Fig. \ref{PID_all_V3} (top panel)
shows the velocity $\beta$ of the particles as a function of their polarity
times momentum for the TOF system. Together with the experimental data,
curves exploiting the relation $p=\beta\gamma m_0 c$, with $m_0$ as
particle rest mass in vacuum and $\gamma=(1-\beta^2)^{-1/2}$,
are shown for different particle species.
The corresponding distribution for the TOFino system is not shown. It has a lower spatial and temporal resolution.

In addition to the arrival time, also the signal height is measured in TOF and TOFino. The latter one is proportional to the energy loss in the
scintillators. The so obtained distribution of the particle energy loss in the
scintillator of the TOF system is shown in Fig.~\ref{PID_all_V3} (TOF $dE/dx$, center panel)
as a function of polarity times momentum. The solid lines are obtained by fitting
the Bethe-Bloch formula \cite{PDG} to the proton distribution and then scaling
the other curves according to the different particle masses.
The separation power of this observable is comparable to the one obtained from the velocity measurement.

The bottom panel of Fig. \ref{PID_all_V3} shows a similar distribution, but for the combined
energy loss in the 4 MDCs of one sector.
The energy-loss information is derived from the sense wire signals after unfolding effects due to the arrival time distribution of the drifting electrons which vary with track incident angle \cite{PhD_Schmah,PhD_Markert}.
The solid curves in the same panel were calculated using the Bethe-Bloch formula.
For the identification of $\pi^\pm$ and protons detected in the TOF,
the $\beta$-versus-momentum information is used and the particles
are selected by applying a graphical cut around the solid lines shown in
Fig.~\ref{PID_all_V3} (top).
The identification of $\pi^\pm $ and protons in the TOFino is performed
in the same way, if the hit multiplicity per paddle is equal to 1;
in case of multiple hits, only the corresponding MDC $dE/dx$ information is used
to perform the selection.

\begin{figure}[!htb]
\begin{center}
\includegraphics[angle=0,scale=0.31]{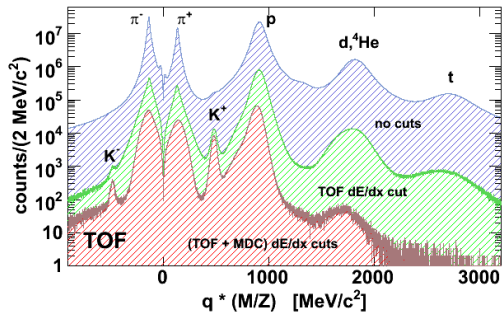}
\caption[]{Color online. Polarity times mass over charge distributions for particle tracks measured in the TOF detector for different cut conditions.
The spectrum on top (blue curve) shows the original distribution,
whereas the lower two spectra show the effects of the graphical cuts
in the TOF $dE/dx$ (green histogram) and additionally the MDC $dE/dx$ (red histogram) distributions, respectively.
\label{Kaon_mass} }
\end{center}
\end{figure}

For the $K^\pm$ identification, only the TOF data are considered,
because of rather large background in the TOFino due to its limited granularity and time
resolution. As evident from Figs.~\ref{PID_all_V3} (top) and \ref{Kaon_mass},
the huge $\pi^\pm$ and proton background hides completely the $K^\pm$ signals.
In order to reduce this background, graphical cuts based on calculated $dE/dx$ distributions have been applied.
The effect of these cuts on the signal-to-background ratio for the ${K^{\pm}}$ signal
is exhibited in Fig.~\ref{Kaon_mass}, where the application of the
TOF $dE/dx$ and the MDC $dE/dx$ vs. momentum selection reduces the background such that
${K^+}$ and ${K^-}$ signals are clearly visible above the background.
\subsection{$\mathbf{K^+}$ and $\mathbf{K^-}$ spectra}
\begin{figure}[!htb]
\begin{center}
\includegraphics[viewport= 35 33 536 665,angle=0,scale=0.37]{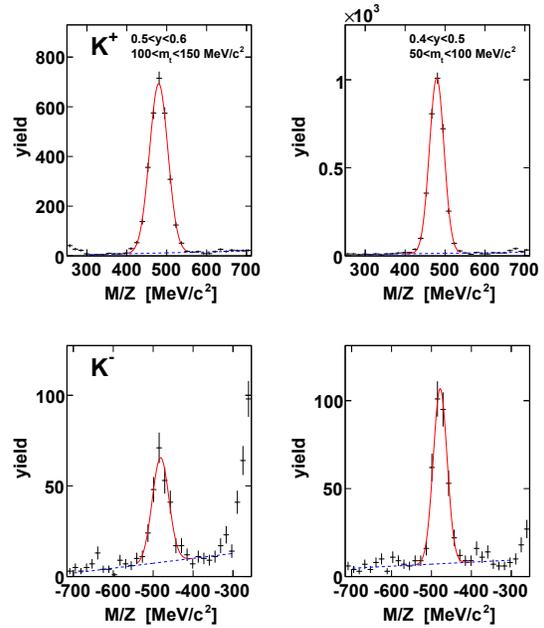}
\caption[]{Reconstructed mass distributions of ${K^+}$ (top) and ${K^-}$ mesons (bottom)
for different phase space regions (left: $0.5 < y < 0.6$ and $100 MeV/c^2 < m_{t} < 150 MeV/c^2$, right: $0.5 < y < 0.6$ and $100 MeV/c^2 < m_{t} < 150 MeV/c^2$). The solid lines represent combined fits to the signal and background,
whereas the dashed lines show the background part only.
} \label{kmass}
\end{center}
\end{figure}

Examples of the mass distribution obtained by applying both the TOF $dE / dx$ and MDC $dE / dx$ cuts are
shown in Fig.~\ref{kmass}. The data are shown in intervals of
rapidity, $y=\frac{1}{2} \ln((E+p_z)/(E-p_z))$, and subtracted transverse mass
($m_{t}$-$m_{K^{\pm}}$ with $m_{t} = \sqrt{M_{K^\pm}^2 + p_t^2}$), where $p_t$, $p_z$ and $E$ are the transverse
and longitudinal momenta and the total energy of the particles, respectively.
The solid lines in Fig.~\ref{kmass} corresponds to a combined fit of the signal and background,
while the dashed lines indicate the background.
A Gaussian fit of the signal after the background subtraction and integrating
over the whole covered phase space gives the following mean values and dispersions for the
reconstructed masses:
$\langle m_{K^\pm}\rangle= 485 \, {\rm MeV/c}^2$, $\sigma_{K^\pm}= 20 \, {\rm MeV/c}^2$.
The slight deviation of about 2\% from the nominal rest mass of $m_K= 494\, {\rm MeV/c}^2$
\cite{PDG} is attributed to imperfections in the time-of-flight calibration, to be improved
for further analysis. Note, however, that all kinetic quantities are derived from the
measured momentum while the time-of-flight serves for particle identification only.
The signal-to-background ratio (S/B) for kaons depends on the location in phase space.
It varies for ${K^+}$ mesons from 1.1-35.6 and for ${K^-}$ mesons from 0.5-4.1.

\begin{figure}[!htb]
\begin{center}
\includegraphics[viewport= 35 30 565 670,angle=0,scale=0.38]{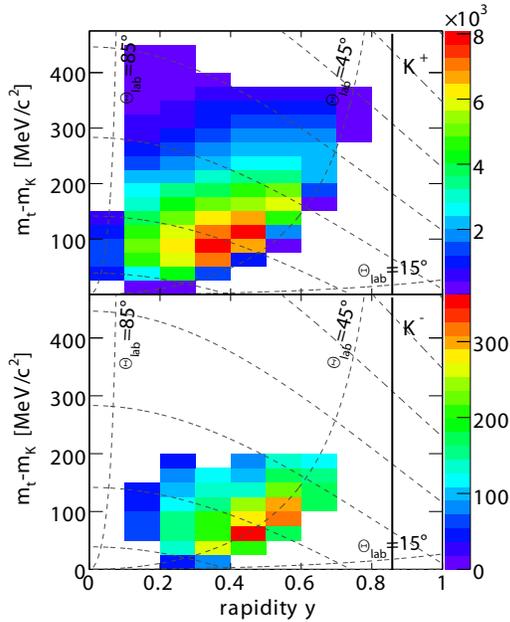}
\caption[]{Color online. Distribution of measured ${K^+}$ (top) and ${K^-}$ (bottom) yields in the TOF detector as a function of the subtracted transverse mass $m_t-m_{K^{\pm}}$ and the laboratory rapidity $y$ of . The color code refers to the counting rate.
The dashed lines indicate the polar angles $\Theta = 15^{\circ}, 45^{\circ}$ and $85^{\circ}$
in the laboratory system. The  dotted curves are for constant kaon momentum
starting from $p=200$\,MeV/c in steps of $200$\,MeV/c.
The vertical solid line depicts mid-rapidity.
\label{kaonspace}}
\end{center}
\end{figure}

The raw phase space distribution for the identified ${K^+}$ and ${K^-}$ mesons
after background subtraction is exhibited in Fig.~\ref{kaonspace},
where the color code refers to the counting rate. Due to low statistics or large correction
factors, mainly connected to the detector acceptance, not all of the shown bins are used in the further data analysis: only the rapidity bins between $0.1<y_{LAB}<0.7$ have been considered. The three dashed lines in Fig. \ref{kaonspace} refer to the laboratory polar angles of
$18^{\circ}, 45^{\circ}$ and $85^{\circ}$, respectively. The $K^\pm$ acceptance
is limited to the TOF region $45^{\circ} < \theta < 85^{\circ}$.
\subsection{Efficiencies}
\begin{figure}[!htb]
\begin{center}
\includegraphics[viewport=20 210 580 800,angle=0,scale=0.44]{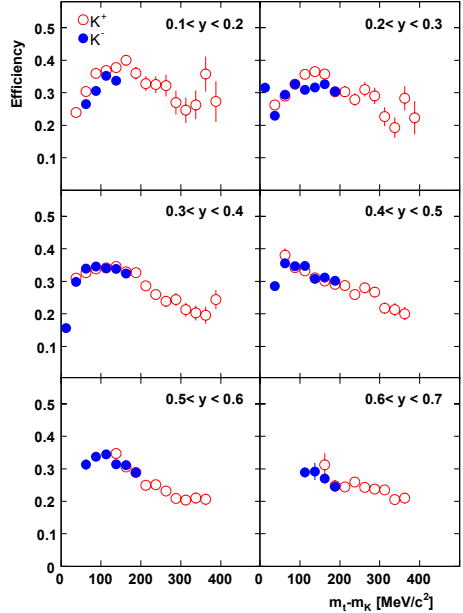}
\caption[]{Color online. Reconstruction efficiency for ${K^+}$ and ${K^-}$
as a function
of the transverse mass for different rapidity intervals.
The acceptances are not included (see text).
\label{effmdc}}
\end{center}
\end{figure}

In order to extract quantitative information on the $K^\pm$ yields and momentum distributions
over the whole phase space, acceptance and efficiency corrections must be applied.
The total yields are finally obtained by extrapolating the measured phase space to $4\pi$.
The geometrical acceptance for ${K^+}$ and ${K^-}$ has been determined
as a function of the emission rapidity and transverse mass.
A full-scale simulation of the HADES spectrometer was performed with the
GEANT3 package \cite{geant}. Kaons and antikaons were generated with a flat (white) distribution in rapidity ($0<y_{LAB}<0.7$) and transverse mass ($0~MeV/c^2<m_t-m_K<450~MeV/c^2$). Then the generated particles were propagated through the detectors and the resulting hits subjected to the selection criteria applied to the real data. The reconstruction efficiency has been evaluated in two steps as a function of $y$ and $m_t-m_{K}$.
First, simulated ${K^+}$ and  ${K^-}$ tracks were embedded in real Ar + KCl
experimental events and the single-track reconstruction efficiency was estimated.
In a second step the PID efficiencies for $K^\pm$ were calculated using
experimental data only, in order to reduce the systematic uncertainty introduced by the
digitization. The efficiency of the TOF $dE / dx$ and MDC $dE / dx$ cuts were
calculated selecting a kaon sample using only one of the two cuts and estimating
the reduction of the signal in the sample when the second cut is applied \cite{Dip_Lorenz}.
This procedure was applied for different rapidity and transverse mass intervals.
The total reconstruction efficiency was obtained by multiplying the three components: track reconstruction, TOF $dE / dx$ cuts and MDC $dE / dx$ cuts.
Assuming that the efficiency for a given momentum slice is independent of the emission angle,
the total efficiency was extrapolated also for the $y - m_t$
bins with low statistics.
The reconstruction efficiency for ${K^+}$ and ${K^-}$ has also been estimated using only simulations, exploiting the embedded tracks method. The digitized data have been tuned such that the simulated $dE / dx$ distributions of both MDC and TOF detectors reproduce the experimental distributions. The so obtained efficiencies show a very good agreement (within  5\%) with the efficiencies extracted from the experimental data and provide the basis for a realistic estimation of the systematic errors.
The resulting efficiency is shown in
Fig.~\ref{effmdc} as a function of $m_t-m_K$ for different rapidity bins.
These values do not include the acceptance which varies from 20-40\%
for both particles analyzed only in the TOF region and in the four sectors
equipped with four MDC planes each. The phase space coverage is displayed in
Fig.~\ref{kaonspace}.

\subsection{${\phi}$ mesons}
\begin{figure}[!htb]
\begin{center}
\includegraphics[angle=0,scale=0.26]{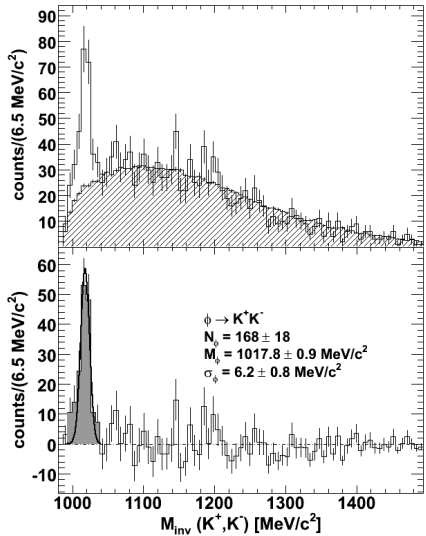}
\caption[]{Invariant-mass distribution of ${K^+}-{K^-}$ pairs (top).
The combinatorial background (shaded area) is obtained by the mixed-event technique.
The background-subtracted distribution (bottom) shows a $\phi$ meson signal
(grey area with a Gaussian fit) with a total yield of $168 \pm 18$ counts.
\label{phi_inv_mass_D}}
\end{center}
\end{figure}

In order to reconstruct $\phi$ mesons, ${K^+}$ and ${K^-}$,
identified in both the TOF and TOFino system and in all six sectors,
are combined to pairs after the application of quality cuts which
are more selective in the case of the TOFino detector.\\
The resulting invariant-mass distribution for the ${K^+}-{K^-}$ pairs is shown in
Fig.~\ref{phi_inv_mass_D} (top) where the $\phi$ signal is clearly visible.
Despite of the fact that the kaon identification is more selective for the particles
which hit the TOF than the TOFino detector, the presence of the peak at the nominal $\phi$ mass with signal to background ratio better than 1 (see Fig. \ref{phi_inv_mass_D}) shows that the purity of our selection is satisfactory for this investigation.
Nevertheless, the continuum visible under the $\phi$ peak is contaminated with misidentified kaons and cannot be considered for quantitative conclusions on the ${K^+}-{K^-}$ non-resonant pair yield.
To extract the $\phi$ signal, the mixed-event technique was employed
for the determination of the combinatorial background. Since a four-fold KCl
target stack was used, only events were combined for which the reaction took place in the
same target segment. The selection was done by calculating the minimal distance of the global event vertex w.r.t. the nominal target positions.
Additionally, only events belonging to the same centrality class ($MUL \pm 4$) were combined. The resulting combinatorial background distribution
is shown by the hatched histogram in Fig.~\ref{phi_inv_mass_D} (top),
where the normalization has been obtained by scaling the mixed-event distribution
to the same-event distribution (arithmetic sum) in the invariant mass region
1050 - 1400~MeV/c$^2$.

Fig. \ref{phi_inv_mass_D} (bottom) shows the signal distribution after the
background subtraction. We obtain for the reconstructed $\phi$ meson mass
$M_{\phi}= (1017.8 \pm 0.9)$\,MeV/c$^2$, a width of $\sigma_{\phi}=(6.2 \pm 0.8)$\,MeV/c$^2$
and a total statistics of $168 \pm 18$ counts.
In order to extract the geometrical acceptance and the reconstruction
efficiency a GEANT3 simulation of a ''white'' $\phi$ meson spectrum has been
propagated through all the analysis steps. Each simulated  ${K^+}-{K^-}$ pair
from a $\phi$ decay
has been embedded in a real event to provide a realistic environment.
The inclusive reconstruction efficiency for the  $\phi$ meson has been evaluated
to be about $1$\%. In addition the fraction of $90$\% of all $\phi$ is lost due to the acceptance and furthermore only 49.2\% of all $\phi$ mesons decay into the channel ${K^+}$ or ${K^-}$, when using the free $\phi$ decay branching ratio.

\section{Experimental results}

\subsection{Transverse mass spectra}

\begin{figure}[!htb]
\begin{center}
\includegraphics[viewport= 170 25 320 520,angle=0,scale=0.39]{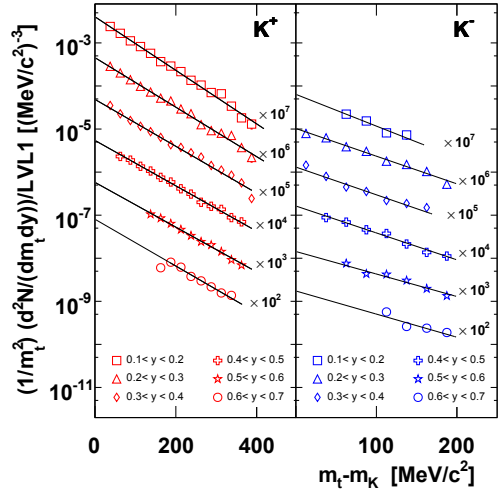}
\caption[]{Color online. Invariant transverse-mass spectra of reconstructed ${K^+}$ (left) and ${K^-}$ (right) mesons.
For the sake of clarity the spectra are plotted for several laboratory rapidity bins,
see legends.
The spectra are scaled by the factors as indicated in the plots.
\label{mtKaons}}
\end{center}
\end{figure}

\begin{figure}[!htb]
\begin{center}
\includegraphics[viewport= 25 40 575 700,scale=0.29]{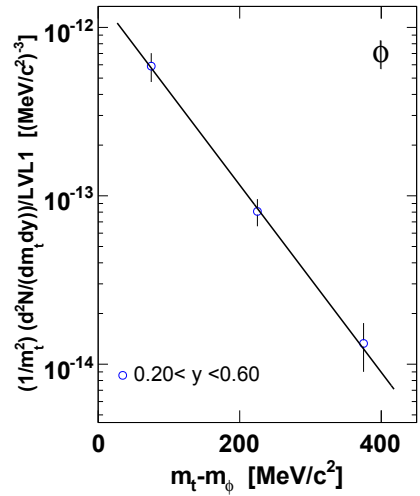}
\caption[]{Invariant transverse-mass spectrum of reconstructed $\phi$ mesons
for the rapidity range $0.2 < y < 0.6$.} \label{mtphi}
\end{center}
\end{figure}

The invariant transverse mass spectra for ${K^+}$, ${K^-}$ and $\phi$ for different laboratory rapidity bins
are exhibited in Figs.~\ref{mtKaons} and \ref{mtphi}, respectively.
These distributions were obtained by applying the efficiency correction to the raw data,
including also the decay probability of each particle in channels that are not
reconstructed in this analysis. The distributions show the number of counts
per LVL1 trigger, per transverse mass and rapidity unit, divided by $m_t^2$.
This representation is chosen to ease the comparison with a Boltzmann distribution.
A fit according to
\begin{equation}
\label{bolz_eqn}
\frac{1}{m_{t}^{2}} \frac{d^2N}{dm_{t}dy} = C(y) \,
\exp \left( -\frac{(m_t-m_0)c^2}{T_B(y)}  \right)
\end{equation}
\begin{figure}[!htb]
\begin{center}
\includegraphics[viewport= 55 35 470 709,scale=0.39]{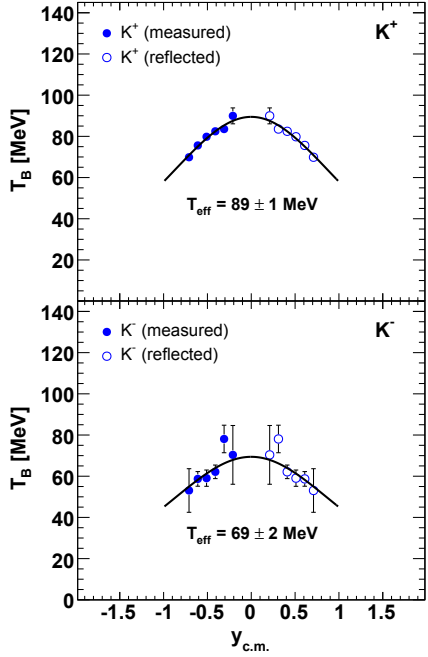}
\caption[]{Rapidity dependence of the inverse slope parameters of ${K^+}$ (top)
and ${K^-}$ mesons (bottom).
The full symbols show the measured data whereas the open ones are reflected data with
respect to the c.m. rapidity. The lines represents fits with Eq.~(\ref{TB_func}).
        } \label{ktb}
\end{center}
\end{figure}
yields the solid straight lines shown in Figs.~\ref{mtKaons} and \ref{mtphi}, which describe indeed the data quite well. The inverse slopes of each distribution, $T_B(y)$, is shown in Fig.~\ref{ktb} as a function of the center-of-mass
(c.m.) rapidity ($y_{c.m.}=y-y(c.m.)$, where $y(c.m.)=0.858$ for symmetric collisions at 1.756 AGeV)
for ${K^+}$ (top) and ${K^-}$ mesons (bottom).
The full symbols display the measured data, whereas the open ones are reflected data with
respect to the c.m. rapidity. The curves in Fig. \ref{ktb} represent fits with
\begin{equation}
\label{TB_func}
T_B(y) = \frac{T_{\mathrm{eff}} } {\cosh(y)},
\end{equation}
assuming a thermal source.
The obtained parameter $T_{\mathrm{eff}}$ represents the inverse slope at mid rapidity
and may be considered as an effective temperature of the kinetic freeze-out of the
respective particle.
In particular, one can see from Fig.~\ref{ktb} that $T_{\mathrm{eff}}$
is lower for ${K^-}$ than for ${K^+}$.
This can be attributed to different freeze-out conditions for the two meson species
\cite{Foerster03}.

\subsection{Rapidity distributions}

\begin{figure}[]
\begin{center}
\includegraphics[viewport= 65 35 470 709,scale=0.4]{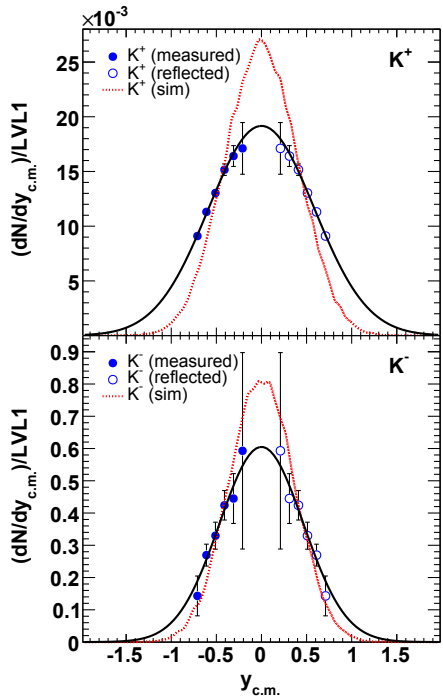}
\caption[]{Rapidity density distributions of ${K^+}$ (top) and ${K^-}$ mesons (bottom).
The full symbols show the measured data whereas the open ones are reflected data with respect to the
c.m. rapidity. The curves represent Gaussian fits to the data. The overlayed dotted lines are simulated isotropic thermal distributions which have been normalized to the data (see text).
\label{krap}}
\end{center}
\end{figure}

The fitted ${K^+}$ and ${K^-}$ invariant transverse-mass distributions $d^2N/(dm_tdy)$ in Fig.~\ref{mtKaons} are integrated within the interval $0 < m_t - m_K < \infty$ in order to obtain the meson yield
per rapidity unit. A different approach to determine this integral is to integrate the measured data points and use the fits only for the extrapolation to the unmeasured phase-space regions. However, the difference in yield to the used method is negligible.
The resulting rapidity density distributions
of ${K^+}$ and ${K^-}$ are displayed in Fig.~\ref{krap}.
The full circles show the values calculated by the integration of
Eq.~(\ref{bolz_eqn}), i.e. by means of
\begin{eqnarray}
\label{dndy_temp}
\left. \frac{dN}{dy} \right|_{y_{i}} & = & C(y_i) \left[ (m_{0} c^2)^{2} T_{B}(y_i)\, + \, \right. \nonumber \\
& &  \left. 2 m_0 c^2 T_{B}^{2}(y_i)+2T_{B}^{3}(y_i) \right].
\end{eqnarray}
The parameters $T_B(y_i)$ and $C(y_i)$ are taken from the fits with Eq.~(\ref{bolz_eqn}) to the invariant transverse mass spectra. The open circles represent the reflection with respect to the center-of-mass rapidity.
The rapidity density distributions have been fitted with Gaussian functions (solid curves represented in the two panels of Fig.~\ref{krap}). The distribution for the ${K^+}$ has a larger width than for the ${K^-}$,
a fact which is also observed in other measurements \cite{Foerster07}.\\
In the fitting procedure, the center of the Gaussian
was fixed at zero, taking into account the symmetry of the reaction.
By integrating the fit function one obtains the kaon multiplicity per triggered event (LVL1).
If the reflected points are included the extrapolation of the measured yield into the unmeasured rapidity region amounts to 35\% and 31\%  of the determined total yield for ${K^+}$ and ${K^-}$, respectively.
The resulting multiplicity values are collected in Tab.~\ref{TempMulti}.
The first error following the multiplicity value indicates the statistical error.
The evaluation of the systematic error is done in two steps. First, the graphical
cuts on the MDC and TOF $dE/dx$ distributions have been varied in width within 30\% and the corresponding variation of the multiplicity calculated. This results in the first systematic error following the statistical one.\\
As shown in Fig. \ref{krap}, the Gaussian parametrization of the data reproduces in a satisfactory way the experimental distribution.
To check the sensitivity of the multiplicity results to the assumed parametrization, a double Gauss fit was also applied to the data following the procedure described in \cite{Alt08}. The resulting uncertainty on the kaon multiplicity is shown by the third error in Tab.~\ref{TempMulti}.
The systematic error of the extracted inverse slope values has been estimated by varying the graphical cuts on the MDC and TOF $dE/dx$ distribution of 30\% as already mentioned above.\\
Figure  \ref{krap} shows in addition to the experimental data and the Gauss fits the rapidity density distributions obtained from simulations of an isotropic thermal distribution of ${K^+}$ and ${K^-}$ respectively. The simulated distributions are represented by the dotted lines in upper and lower panels of Fig. \ref{krap} and is normalized to the total integral of the experimental ones.
The measured  distributions are systematically wider than the simulated ones (see Tab.~\ref{TempMulti}). For $K^+$ the increase in width is larger and amounts to roughly 30\%.\\
Deviations from a thermal picture of a particle emitting source are due to anisotropies of the angular and/or of the momentum distributions. The limited acceptance of the HADES detector precludes such a differentiation at the current stage of analysis.

The rapidity density distribution of the measured $\phi$ mesons contains only one data point.
Therefore,  one has to fix the width of the distribution in order to estimate
the $\phi$ multiplicity. If a thermal description of the rapidity density distribution is assumed,
 its width $\sigma_y$ can be calculated according to 
\begin{equation}
\sigma_y = \sqrt{\frac{T_{\mathrm{eff}}} {m_\phi c^2}},
\label{phi_width}
\end{equation}
where $T_{\mathrm{eff}}$ represents the effective inverse slope for $\phi$ mesons.
This parameter is calculated taking the inverse slope $T_B$ obtained in fitting
the invariant transverse mass distribution of the $\phi$ meson for the rapidity bin shown in Fig.~\ref{mtphi}
and substituting this value into Eq.~(\ref{TB_func}).

Table ~\ref{TempMulti} shows the total $\phi$ meson multiplicity per triggered event (LVL1) obtained by integrating the so-obtained Gaussian curve. The first error shown is again purely statistical.\\
Due to the extended rapidity bin chosen for the $\phi$ mesons the change of the yield
within the bin is not negligible. Instead of using the center rapidity of this bin we take a value
weighted with the expected thermal distribution across the bin width. This
procedure reduces the extrapolated total yield by around 24\% compared to an extrapolation where the rapidity point is in the center of the bin.
The systematical error estimated for the total $\phi$ multiplicity/LVL1 shown in Tab.~\ref{TempMulti} is obtained by varying the temperature $T_{\mathrm{eff}}$ within its statistical error.\\
Existing theoretical calculations assume an isotropic emission of the $\phi$ meson \cite{BK_Kotte,Zetenyi,Barz}, but since the angular distribution of the emitted $\phi$ has not yet been measured at these energies,
an additional source of systematical error must be considered.
On the base of the before mentioned maximal deviation of the  ${K^+}$ rapidity distribution from an isotropic source,  we assume 30\% to be the maximum error in the evaluation of the width of the rapidity distribution for the $\phi$ meson.
This translates into a second systematic error for the $\phi$ multiplicity/LVL1 which is also presented in Tab.~\ref{TempMulti}. This has been calculated assuming a larger width of the $\phi$ rapidity density function and hence leads to an asymmetric error.\\
The value of the effective temperature of the $\phi$ meson is shown as well in Tab.~\ref{TempMulti} together with its statistical error.

\begin{table*}[hbt]
  \begin{center}
    \caption{Total multiplicity/LVL1 and effective inverse slope parameters $T_{\mathrm{eff}}$
    for ${K^+}$, ${K^-}$ and $\phi$. The first error refers to statistical uncertainties,
    while the second and third errors represent the systematic ones, as described in the text. The fitted Gaussian width of the rapidity density distributions ($\sigma_{exp}$) as well as the calculated width for a thermal and isotropic distribution ($\sigma_{therm}$) of the corresponding effective temperature $T_{\mathrm{eff}}$ are also listed with systematic and statistical errors.\\ }
          \label{TempMulti}
    \begin{tabular}{|c|c|c|c|c|}
    \hline
      particle& multiplicity/LVL1 & $\sigma_{exp}$ $[y_{c.m.}]$& $\sigma_{therm}$ $[y_{c.m.}]$ & $T_{\mathrm{eff}}$ \\ \hline
      \hline
      ${K^-}$&(7.1$\pm$1.5$\pm$0.3$\pm$0.1)$\cdot10^{-4}$ & $0.465 \pm 0.0565$ & $0.374 \pm 0.024$ & $69\pm2\pm4$ \\
      \hline
      ${K^+}$& (2.8$\pm$0.2$\pm$0.1$\pm$0.1$)\cdot10^{-2}$ & $0.586 \pm 0.0223$ & $0.425 \pm 0.011$ &$89\pm1\pm2$ \\
      \hline
      $\phi$& $(2.6\pm0.7\pm 0.1^{+0.0}_{-0.3})\cdot10^{-4}$ & - & $0.287\pm 0.027$  &$84\pm8$ \\
      \hline
    \end{tabular}
  \end{center}
\end{table*}
\section{Discussion}
\begin{figure}[]
\begin{center}
\includegraphics[viewport= 40 30 580 530,angle=0,scale=0.3]{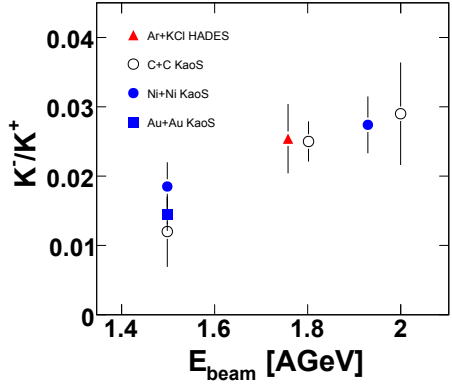}
\caption[]{${K^-/K^+}$ ratios as a function of the kinetic beam energy
for various systems.
The present HADES data point is depicted by the full triangle.
KaoS data are from \cite{Foerster07}.
\label{compKaos2}}
\end{center}
\end{figure}

Let us first compare our data with results recently published by the KaoS collaboration
\cite{Foerster07}.
If the ${K^-/K^+}$ multiplicity ratio is plotted as a function of the beam energy for
the inclusive data and for various systems measured by KaoS \cite{Foerster04,Foerster07}, as shown in Fig.~\ref{compKaos2}, one can see that
 the ratio increases with beam energy and that the HADES data point fits nicely into the energy systematics.

\begin{figure}[]
\begin{center}
\includegraphics[viewport=  120 20 500 500,angle=0,scale=0.3]{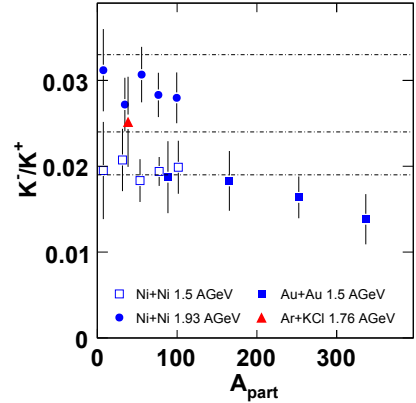}
\caption[]{ $A_{\mathrm{part}}$ dependence of the ${K^-/K^+}$ ratio
for different beam energies and colliding systems. The full triangle depicts the present HADES data.
The other symbols display KaoS results from \cite{Foerster07}.
The dashed lines show the prediction of a thermal model \cite{Cley04,RedlichPrivComm}
for kinetic beam energies of 1.5, 1.76 and 1.93 AGeV (from bottom to top).
\label{compKaos1}}
\end{center}
\end{figure}

In Fig.~\ref{compKaos1} the centrality dependence of the ${K^-/K^+}$ ratio is exhibited
for the Au+Au and Ni+Ni systems and for two different beam energies.
As a measure of the centrality, the number of participants $A_{{part}}$ is chosen.
This quantity is calculated within a geometrical model of interpenetrating spheres where the
impact parameter range is derived from a comparison of the
charged particle multiplicities, as measured in the TOF/TOFino detectors, to that provided by
GEANT predictions with the UrQMD transport model \cite{UrQMD} as event generator.
The full triangle represents the data point measured for Ar+KCl at 1.756\,AGeV in the present
experiment. For the LVL1 trigger we attribute to our data a value of  $A_{{part}} = 38.5 \pm 4$,
being about twice the minimum bias value.
Our data point fits well into the systematics when taking into account the beam energy dependence.
Looking at the empty squares (Ni+Ni at 1.5 AGeV) and full squares (Au+Au at 1.5 AGeV),
it can be seen that the ratio stays fairly constant with $A_{{part}}$, supporting our implicit assumption
on the centrality independence in using
an averaged value of $A_{{part}}$. Together with the data, ${K^-/K^+}$ ratios predicted by the thermal model \cite{Cley04,RedlichPrivComm} are shown for kinetic beam energies of 1.5, 1.76 and 1.93 AGeV. These calculated values are found to be in fair agreement with the experimental values.
The fact that the ratio ${K^-/K^+}$ does not depend noticeably on the centrality of the reaction and that the meson yield is consistent with the prediction
of thermal models has been used in the past to support the hypothesis
that most of the ${K^-}$ are produced in multistep processes like
\begin{eqnarray}
&& N N  \rightarrow  Y K^+ N, \quad E_{th}=1.58\,{\rm GeV},\\
&& Y \pi  \leftrightarrow  N K^-, \label{strExch}
\end{eqnarray}
where $Y$ stands for the hyperons $\Lambda$ and $\Sigma$,
and $E_{th}$ denotes the beam energy threshold in free nucleon-nucleon collisions.
The reaction (\ref{strExch}) is called strangeness exchange.
The dependence of the ${K^+}$ and ${K^-}$ multiplicities
on $A_{{part}}$ has been found to be quite smooth \cite{Foerster07}:
$M_{K^+}\propto A_{{part}}^{\alpha}$ with $\alpha_{K^+}(Ni)=1.26\pm0.06$ and
$M_{K^-}\propto A_{{part}}^{\alpha}$ with $\alpha_{K^-}(Ni)=1.25\pm0.12$.
(In contrast, the $\pi^\pm$ multiplicity is found to linearly dependent on $A_{{part}}$, $\alpha = 1$).
Along these considerations
it has been claimed in \cite{Cley04} that reaction (\ref{strExch})
is the dominant channel for ${K^-}$ production.

Further possible channels for the production of ${K^-}$ in nucleon-nucleon collisions are
\begin{eqnarray}
NN  &\rightarrow& \phi NN \nonumber \\
    &\rightarrow& K^+ K^- NN, \quad E_{th}=2.59\,{\rm GeV}, \label{othC1}\\
NN  &\rightarrow& K^+ K^- NN, \quad E_{th}=2.49\,{\rm GeV}, \label{othC2}
\end{eqnarray}
with thresholds being higher than for the previous reaction.\\
The threshold for production induced by a $\Delta$ resonance $N\Delta \rightarrow K^+ K^- NN$ depends on
the mass (energy) of the resonance, but it is comparable to
the value quoted in Eqs.~(\ref{othC1}) and (\ref{othC2}). Further secondary collisions have been considered in theory \cite{BK_Kotte, Zetenyi,Barz} to contribute to the $\phi$ production such as $\pi N \rightarrow \phi N$, $\pi N(1520) \rightarrow \phi N$, $\rho N \rightarrow \phi N$, $\rho \Delta \rightarrow \phi N$, $\pi \rho \rightarrow \phi $.
It has to be pointed out \cite{BK_Kotte, Zetenyi,Barz} that the predicted $\phi$ multiplicity in sub-threshold heavy-ion collisions, calculated including all the channels mentioned above, stay below the number extracted from the experimental data \cite{Mangiaro03}.

The newly measured HADES data provide for the first time at SIS energies a consistent measurement of ${K^+}$, ${K^-}$ and $\phi$ in the same data sample.
The $\phi/{K^-}$ ratio measured with the HADES spectrometer for Ar+KCl at 1.756 AGeV
is found to be $0.37\pm0.13$, see Tab.~\ref{TempMulti}.
This value translates into a fraction of
$18 \pm 7\%$ of ${K^-}$ coming from $\phi$ decay. This implies that
also below threshold the $\phi$ production contributes significantly to the ${K^-}$ rate,
as already argued in \cite{BK_Kotte}.
Taking for the $\phi$ meson the decay path $c\tau\approx46$ fm as follows
from the vacuum width, one can estimate a fraction of
about 80\% decaying outside the collision zone, for which we employ a radius of
about 10 fm. The question of the in-medium width of the $\phi$ meson is certainly far from being settled, since the few available data do not yet deliver a consistent picture \cite{Metag07}. On the other hand, if we assume that a large amount of $\phi\rightarrow{K^-}{K^+}$ decays would take place in the fireball volume, due to the strong interaction experienced by kaons in medium the resulting signal of in the ${K^-}{K^+}$ invariant-mass spectrum would be washed out. Hence, the ${K^-}$ stemming from $\phi$ decays reconstructed in this analysis are produced mostly outside the medium. This effect will also dilute any observed medium effects of the ${K^-}$.
In order to extract precise information on the equation of state \cite{Ko96,Li97,Cassing97} and on the ${K^-}$-nucleon potential,
one should improve the statistics of such independent measurements for ${K^-}$ and $\phi$, i.e. tag the ${K^-}$ according to their production mechanism. This kind of systematic measurements should be done for different systems and different beam energies.

Also the non-resonant ${K^+}-{K^-}$ production (say by the reaction Eq. (\ref{othC2})) plays in p+p reactions an important role, contributing to about $50\%$ of the overall ${K^+}-{K^-}$  yield \cite{ANKE08}. This contribution is not yet known in heavy-ion collisions, since the ${K^+}-{K^-}$ invariant mass spectrum measured by HADES still contains a non-negligible contribution of fake candidates (see discussion of Fig. (\ref{mtphi})). If one assumes that the contribution of the non resonant ${K^+}-{K^-}$ production in Ar+KCl at 1.756 GeV is comparable with the value measured for p+p at 2.7 GeV  \cite{ANKE08}, taking the ratio $\phi/{K^-}$ measured by HADES, it can be estimated that about 38\% of all produced ${K^-}$ do not stem from strangeness exchange reactions but from the sum of resonant and non-resonant  ${K^+}-{K^-}$ production  (Eqs. \ref{othC1}, \ref{othC2}).
\begin{figure}[]
\begin{center}
\includegraphics[viewport= 54 181 550 670,angle=0,scale=0.39]{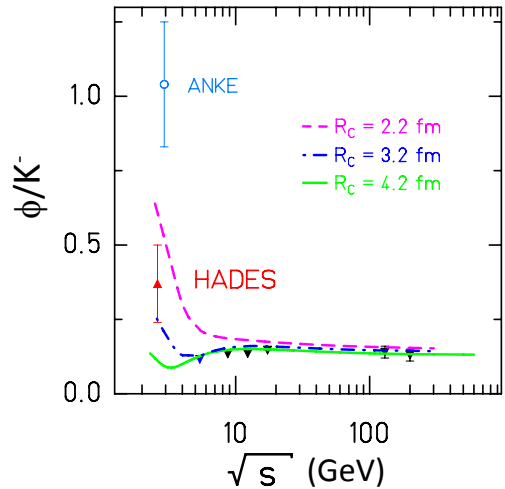}
\caption[]{$\phi/{K^-}$ ratio as a function of center-of-mass energy $\sqrt{s}$. The full triangle on the left shows the HADES point while the other data points represented by full triangles refer to results from heavy-ion measurements at higher incident energies \cite{Ada05,Hol02,Afa00}. The empty circle presents the value measured in p+p at 2.7 GeV \cite{ANKE08}. The lines are predictions of a statistical model \cite{RedlichPrivComm} for three parameters of the correlation radius $R_C$.
\label{ratiophi}}
\end{center}
\end{figure}

This consideration further deprives the strangeness exchange mechanism of its leading role in the ${K^-}$
production in sub-threshold heavy-ion reactions.

In Fig. \ref{ratiophi}, the $\phi/{K^-}$ ratio measured by HADES is compared to the ratios measured at higher energies in Pb+Pb \cite{Afa00} and Au+Au collisions \cite{Ada05,Hol02} and in p+p reactions at 2.7 GeV \cite{ANKE08}. For heavy ion collisions one can see that, while the ratio is rather constant at high energies, it is substantially larger in case of the HADES measurement.

The various curves in Fig. \ref{ratiophi} show the prediction by a statistical model for three different values of the correlation length $R_C$ \cite{RedlichPrivComm}. This parameter is the radius of the volume inside which strangeness is locally conserved in the calculation. One can see that a value of $R_C\leq 3.2 $ fm is needed to fit the prediction by the statistical model to the HADES point.
A smaller value of $R_C$ translates into a reduced volume for local strangeness production and conservation. The fact that the HADES result is consistent with a smaller volume could be interpreted as a hint that the contribution to ${K^-}$ production by strangeness exchange, which is more probable for larger volumes, is diminished in the SIS energy regime.\\
The $\phi/{K^-}$ ratio measured in elementary reactions is thought to be consistent with  $R_C\approx 1 $ fm. The fact that the ratio measured by HADES is smaller than the one extracted for p+p reactions tight above threshold shows the role played by secondary processes in the ${K^-}$ production in sub-threshold heavy-ion collisions. On the other hand, if the strangeness exchange process would be the dominant mechanism for ${K^-}$ production, the $\phi/{K^-}$ ratio would be much smaller than the measured value.

\section{Summary}
In summary,
we report on a first measurement of charged kaons and the  $\phi$ meson production in the same experiment at SIS energies.
The HADES data on ${K^{\pm}}$ fit well into the systematics of the results published by
the KaoS collaboration.
The ${K^-/K^+}$ ratio is consistent with the value predicted by the thermal model.
The effective inverse-slope parameters of ${K^+}$ and ${K^-}$ are determined;
the inverse slope of ${K^+}$ mesons is found to be higher than that for ${K^-}$,
in agreement with the KaoS data.

The $\phi/{K^-}$ ratio has been deduced from our data with improved accuracy. The HADES value is found to be $0.37\pm0.13$,
translating into a fraction of $18 \pm 7$$\%$ of ${K^-}$
stemming  from $\phi$ decays. Since the $\phi$ decays that can be measured via the invariant-mass reconstruction of ${K^+}-{K^-}$ pairs are essentially those happening outside the nuclear medium, this value may be considered as a lower limit.
On the other hand, also the non resonant ${K^+}-{K^-}$ production may play an important role for the total ${K^-}$ yield, but in order to draw quantitative conclusions the purity of the kaon identification in heavy-ion reactions must be improved.
In general it can be concluded that for heavy-ion collisions
at beam energies below the free $\phi$ production threshold, the ${K^-}$ production
cannot be explained exclusively by the strangeness exchange mechanism.\\
Furthermore, the $\phi/{K^-}$ ratio for the Ar+KCl system has been compared to
the ratio measured in heavy-ion collisions at higher energies and to the ratio extracted from p+p reactions at 2.7 GeV. The $\phi/{K^-}$ ratio measured by HADES has been found to be higher than the almost constant value found at higher energies. A correlation parameter  $R_C\leq 3.2$ fm has to be used in the calculation by the Statistical Model to reproduce the HADES measurement. This value is lower than the one used to fit the $\phi/{K^-}$ measured in heavy-ion collisions at higher energies. A reduced value of $R_C$ can be interpreted as a reduced volume in which strangeness production and conservation is taking place and hence to a less significant contribution of the strangeness exchange mechanism, to the ${K^-}$ production, which is more probable for larger volumes. \newline
The contribution of the different secondary processes to the total ${K^-}$ is visible in the difference between the $\phi/{K^-}$ ratio measured in p+p reactions at 2.7 GeV and in Ar+KCl collisions at 1.756 AGeV. The observed relative high ratio measured by HADES seems to deprive the strangeness exchange mechanism from its dominant contribution to ${K^-}$ production, as stressed above.

It is clear that, in order to extract an excitation function, further systematic measurements of the $\phi$ and ${K^-}$ production
have to be performed. HADES is a suitable apparatus for this kind of studies, especially due its capability to reconstruct the $\phi$ not only via the charged kaon decay but also via the direct electromagnetic decay into lepton pairs. Due to the fact that leptons do not undergo strong interactions and therefore are able to leave the reaction zone nearly undisturbed, one will be able to distinguish between those $\phi$ mesons which decay inside the medium and those who decay outside. This will provide an observable directly sensitive to the predicted medium modifications.

\subsection*{Acknowledgements}
We gratefully acknowledge the useful discussions with H. Oeschler and K. Redlich. In particular we thank K. Redlich for providing the calculations by the statistical model.
The HADES collaboration gratefully
acknowledges the support by BMBF grants 06TM970I, 06GI146I, 06F-140,
and 06DR135 (Germany), by GSI (TM-FR1, GI/ME3, OF/STR), by Excellence
Cluster of Universe (Germany), by grants GA
AS CR IAA100480803 and MSMT LC 07050 (Czech Republic), by grant KBN
5P03B 140 20 (Poland), by INFN (Italy), by CNRS/IN2P3 (France), by
grants MCYT FPA2000-2041-C02-02 and XUGA PGID T02PXIC20605PN
(Spain), by grant UCY-10.3.11.12 (Cyp\-rus), by INTAS grant
06-1000012-8861 and EU contract RII3-CT-2004-506078.

\end{document}